# Community Action on FAIR Data will Fuel a Revolution in Materials Research

**Authors**: LC Brinson, LM Bartolo, B Blaiszik, D Elbert, I Foster, A Strachan, PW Voorhees
*(Complete author affiliations and contact information at end)*

*Widely shared and accessible materials data are the key to a world in which accelerated materials development addresses society's greatest challenges. We present a roadmap for connected materials data to enable researchers, designers, and manufacturers to harness its power*.

Little of the data – arguably the most important product of worldwide materials research worldwide – are shared in forms usable by others. The small and biased proportion of results published are buried in plots and text licensed by journals. This situation wastes resources, hinders innovation, and, in the current era of data-driven discovery, is no longer tenable. In this Perspective we propose specific synergistic, collaborative, and global actions to enable the assembly of large quantities of **FAIR** (**F**indable, **A**ccessible, **I**nteroperable, **R**eusable) (1) materials data. We provide a *context* to comprehend what FAIR data can mean for materials scientists, a *motivation* for the adoption of FAIR principles, and a *perspective* on how widespread adoption of FAIR data can advance their science.

A decade ago, the U.S. Materials Genome Initiative (MGI) (2) articulated goals of accelerated materials development and deployment via advanced computational methods, integrated and high throughput experiments, with a focus on data standards, sharing, transparency, modeling, and design; a *2021 Materials Genome Initiative Strategic Plan* (2) expands the MGI's scope to encompass a new "Materials Innovation Infrastructure", a focus on AI, and a community network for standards, education and training. Parallel initiatives worldwide are pursuing similar visions (3-7). In Germany, the National Research Data Infrastructure (NFDI)'s MatWerk 2021 has awarded five years of funding to support efforts in FAIR DATA and shared data space for Materials Science and Engineering. In 2021, the UK launched its Innovation Strategy with support for advanced materials & manufacturing, and in 2020 the EU established the OntoCommons for shared materials and manufacturing data ontologies. Japan's Strategic Innovations Program (SIP) created the Design System of Structural Materials in 2020. Such efforts, the rapidly growing number of papers in materials science using machine learning and their citation rates (Fig 4 in (8), (9)), and the emergence of journal publications focused on scientific data and associated metadata (e.g. Nature Scientific Data) make clear the global importance of data to materials science and engineering (10-17).

Yet despite large investments in materials science and engineering – more than $37B in 2018 by US industry alone (18) – most data languish in local storage systems or reports and papers (2, 12, 13). In contrast, imagine being able to "*google"* all materials ever synthesized or predicted, to find organized, annotated, quantitative, referenced, citable, and downloadable data for the subset of materials that have a desired combination of properties and characteristics. Joining MGI and FAIR data brings this vision into reach.

**FAIR (Findable, Accessible, Interoperable, Reusable) materials data**

The FAIR Principles, applicable to *any* type of data, provide unifying guidelines for the effective sharing, discovery, and reuse of digital resources, including data, metadata, protocols, workflows, and software. FAIR data for Materials will enable better science via reproducibility and transparency and provide a path to reward valued data generators. Widespread FAIR data will unleash an era of materials informatics where exploring prior work is nearly instantaneous and drive development of advanced analytics and machine learning for materials.

Realizing the promises of MGI and FAIR, however, requires community agreement and implementation. General FAIR principles (1) are necessary but not sufficient to transform the field of materials, where varied interpretations and definitions of basic composition and property terms hold back effective implementation (19). Each data type has different forms, vocabularies, and descriptors across material types, from polymeric systems to metals, biomaterials, ceramics, and functional materials.

As depicted in **Figure 1**, making materials data FAIR need not involve heroic efforts but does require attention and deliberate and consistent adoption of available protocols. For example, the use of globally unique, persistent identifiers (UUIDs or PIDs) as long-lasting references for digital resources is "FAIR", while the typical protocol of making data "available upon request" is "not FAIR".

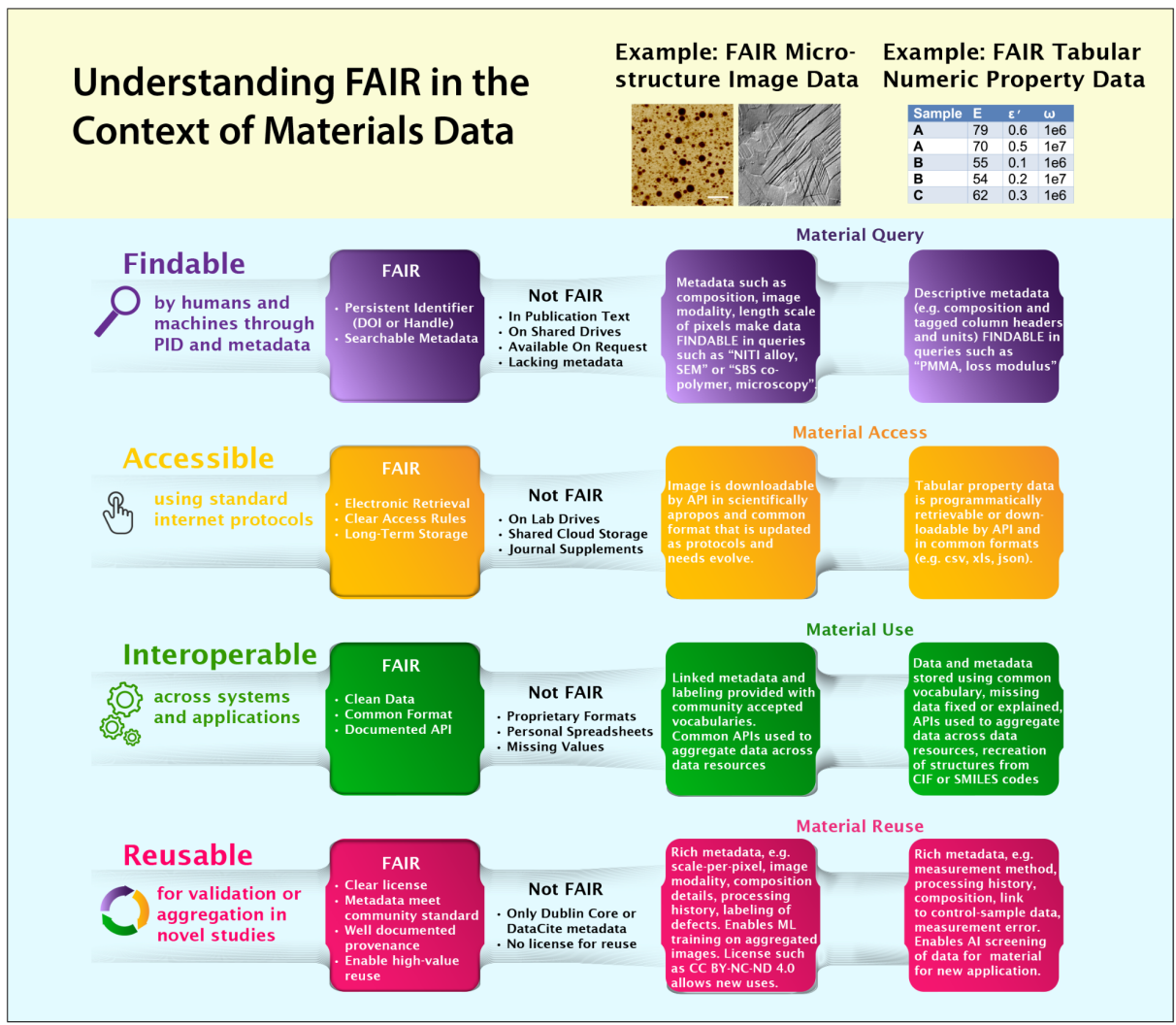

*Figure 1.* Concise definitions of the FAIR principles translated to two specific examples of materials research datasets: 1) microstructure images, containing nuances such as grain boundaries, dislocations, inclusions, and/or dispersion of particles; 2) numeric data representing spectral responses for a property value across space, temperature, time, and frequency. The inherently heterogeneous data include image and video data at many length scales; scalar, vector, tensorial, and tabular physical property values; text and numbers defining compositions and processing conditions; and metadata on computational methods, experimental protocols, assumptions, and analyses. Findable and Accessible materials data resources include the Materials Project, OpenKim, Materials Data Facility (MDF). Interoperability is being tackled by the Crystallographic Information Framework (CIF, 20), Simplified Molecular-Input Line-Entry System (SMILES, 21), and OPTIMADE (22). Reusability, the ultimate goal of FAIR for materials science, and depends on development and consistent use of metadata standards.

**Materials Data Stakeholders: Barriers and Hopes**
In planning the operational and cultural changes required to achieve broadly FAIR materials data, we must consider the agendas, needs, and concerns of five large cadres of stakeholders: *researchers* who generate data; *developers of hardware and software tools* used to produce research results; *publishers* and *repository developers* that transmit research results; *funders* who support research; and *consumers* who use data. We interviewed members of each group in developing our recommendations.

The number one **barrier** to FAIR materials data is fear of productive *time* lost in archiving, cleaning, annotating, and storing data and associated metadata. Funders and researchers are concerned about lost productivity, publishers about barriers and delays to publication when data sharing is enforced, and consumers about spending time finding data in a new and unfamiliar landscape. Other major concerns identified include: *navigation of licensing*, *fear of being scooped / fear of losing credit*, *intellectual property restrictions for materials data*, and *quality control for data housed in repositories*.

Stakeholders simultaneously expressed **great hope** for a data-rich future where *journal articles are linked with FAIR datasets*; ever-growing supplementary information (SI) is replaced with references to *cleanly annotated data in repositories*; *measures of quality and FAIR metrics* naturally evolve for housed data; and data are *citable, findable, and reusable, and have significantly larger impact*.

Achieving widespread FAIR materials data requires overcoming both sociological and technical challenges. To combat the major fear of "lost time", we need demonstrations of FAIR data enabling success, incentives for sharing FAIR data, and infrastructure to simplify or automate data upload and annotation. Data literacy and best practices need to become part of education and researchers' daily workflow so that making data FAIR is no longer a taxing after-thought nor a fear of lost credit. Sustainability models must be developed and implemented to support hosting large quantities of data and required infrastructure.

**A Roadmap to FAIR Materials Data Infrastructure**
We depict in a roadmap (**Figure 2**) both individual and community-level actions to accelerate materials research via FAIR data. The community-level actions are:

- **Incentivize and recognize data literacy and reward best practices in data stewardship.** Track *"data use" citations* and create a *data citation index* to reward publishing of FAIR data; create open educational content for FAIR materials data methodologies.
- **Prioritize capture of materials research products beyond datasets:** Archive post-processing methods, trained models, and codes; establish links between materials data repositories and associated models/software.
- **Establish benchmark materials datasets** of high-value and high-profile to drive algorithm development. Establish an award for materials discoveries based on prior data.
- **Define high-impact community data generation tasks in subfields of materials science.** Challenge materials subfields to prioritize specific data products (e.g., microstructural image collections) for transformational change. Engage repositories and communities to catalyze these changes.
- **Promote trustworthy repositories.** Define audit and certification criteria for materials repositories to ensure long-term storage, access, and preservation of data as part of the global materials data infrastructure.

- **Collect and publicize success stories.** Collate compelling examples of data-driven approaches used to advance materials research, curated and promoted by professional organizations and funding agencies.

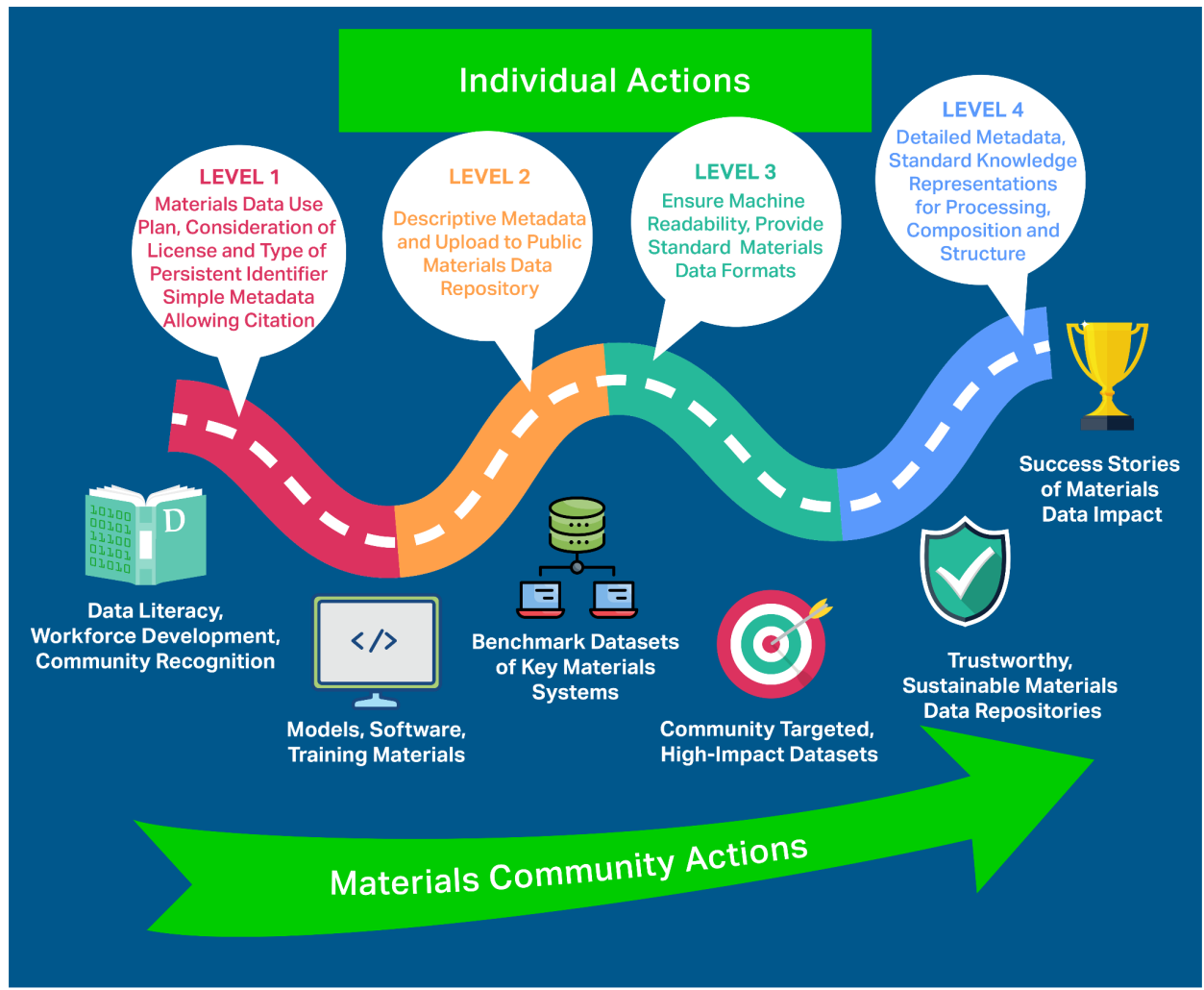

*Figure 2. Roadmap towards FAIR materials data, including four levels of individual action built on a foundation of community actions, all of which create value and motivate change to accelerate the widespread adoption of FAIR in the materials domain. Community networks such as MaRDA (Materials Research Data Alliance) play a vital role in enhancing synergy between individual and community driven actions, including by synchronization and promotion of efforts and defining and updating a formal roadmap with tangible goals and target dates.*

**Figure 2** also shows four levels of individual action that can be taken by researchers, research groups, and labs to produce FAIR data and enhance scholarly output. The practices encompassed by these levels – organized in roughly increasing order of complexity – can be adopted one at a time, in various orders, and in any materials research effort. In each Level, actions are labeled with F, A, I, or R:

*Level 1: Planning and Preliminary Data Submission*

Define materials data and metadata at project outset. Consider how the data could be reused by others for tasks unrelated to the originator's work, quantifying and capturing uncertainties is often critical in this step. (R) Use electronic lab notebooks to facilitate data and metadata extraction as well as documenting and publishing data management workflows (23). (I) Make data available through a general repository with persistent identifiers (e.g., DOIs) for datasets (e.g., Zenodo, Figshare, Dryad). (F) Include licensing information and *how to cite* examples in metadata, as supported by Figshare, Dryad, MDF, and nanoHUB. (R)

*Level 2: Materials-Specific Metadata and Complete Submission*
Include detailed descriptive metadata, via for example metadata columns in a CSV data file. (R, F) Place data and metadata in materials-specific repository (F, A) with fields designed to handle and share materials relevant terms: e.g., OpenKIM for interatomic models, MDF for heterogenous datasets up to many terabytes in size, Foundry for structured ML-ready datasets, MaterialsMine for polymer nanocomposites and structural metamaterials, or AFLOW or OQMD for DFT calculated data on thermodynamic properties of crystallographic materials.

*Level 3: Enhanced Functionality*
Ensure data and metadata are both human and machine readable; employ "tidy" data protocols (24). Place data in repositories that support long-term storage and query via standard interfaces interfaces (e.g., APIs) (F, A): e.g., Materials Project, AFLOW, OQMD, MDF.

*Level 4: Community standards, Provenance, and Reusing Data*
Use community standards for knowledge representation and standard file formats for data and metadata. Examples include SMILES for molecules and CIF for crystals which can be automatically processed by visualization and machine learning packages. (I) Include metadata that points to other metadata as needed to provide detailed context, ensure software and protocols have well-defined and verified requirements (inputs) and services (outputs). (I) Reuse others' data in your research, e.g., for benchmarking or in analyses to create new data. (R)

Community networks such as the US Materials Research Data Alliance (MaRDA) and materials subgroups in the Research Data Alliance (RDA), working closely with stakeholders, can support the transition to FAIR materials data. Critical actions include providing the coordination and engagement required to develop and maintain protocols, standards and best practices; development and promotion of sustainability models for materials data repositories; regular updates to the roadmap to FAIR materials data and annual scoring of the communities' progress.

New data-driven approaches to materials innovation promise transformational contributions to human health and prosperity, but are hindered by inadequate access to data on materials and material properties. The roadmap presented here highlights policies and practices that the materials community and individuals can adopt to catalyze the creation of a distributed, yet unified, worldwide materials innovation network within which data can be reused and recombined to unleash a new era of accelerated innovation and progress.

**Author list with complete affiliation information:**

L. Catherine Brinson, Sharon C and Harold L Yoh III Distinguished Professor, Donald M Alstadt Department Chair, Department of Mechanical Engineering and Materials Science, Duke University, Durham NC 27708, cate.brinson@duke.edu

Laura M. Bartolo, Senior Research Associate, Coordinator, Data and Databases, Center for Hierarchical Materials Design, Northwestern University, Evanston IL 60201, laura.bartolo@northwestern.edu

Ben Blaiszik, Research Scientist, Globus, University of Chicago, Chicago, IL 60637; Data Science and Learning Division, Argonne National Laboratory, Lemont, IL 60439; Materials Data Facility, blaiszik@uchicago.edu

David Elbert, Research Scientist, Hopkins Extreme Materials Institute and Chief Data Officer, PARADIM Materials Innovation Platform, Johns Hopkins University, Baltimore, MD, 21218, elbert@jhu.edu

Ian Foster, Arthur Holly Compton Distinguished Service Professor, Department of Computer Science, University of Chicago; Distinguished Fellow, Senior Scientist, and Director, Data Science and Learning Division, Argonne National Lab, foster@uchicago.edu

Alejandro Strachan, Professor of Materials Engineering, Deputy Director, nanoHUB, Purdue University, West Lafayette, IN 47907, strachan@purdue.edu

Peter W. Voorhees, Frank C. Engelhart Professor of Materials Science and Engineering, Director Center for Hierarchical Materials Design, co-Director Northwestern Argonne Institute for Science and Engineering, Chair Department of Materials Science and Engineering, Northwestern University, Evanston, IL 60201, p-voorhees@northwestern.edu

**Acknowledgments**

**People**: The authors wish to thank John Allison, Apurva Mehta, Eileen De Guire, Erik Schultes, Daniella Lowenberg, James Warren, Jack Brook and Laura Franklin for helpful conversations as this comment was prepared.

**Funding**: LCB acknowledges support from DOE DE-SC0021358 and NSF CSSI-1835677. PV acknowledges the financial support of award 70NANB14H012 from the U.S. Department of Commerce, National Institute of Standards and Technology as part of the Center for Hierarchical Material Design (CHiMaD). DCE acknowledges support from the National Science Foundation (Platform for the Accelerated Realization, Analysis, and Discovery of Interface Materials (PARADIM)) under Cooperative Agreement DMR-2039380 and DTRA award HDTRA1-20-2-0001. AS acknowledges support from the US National Science Foundation DMREF program (DMREF-1922316) and Network for Computational Nanotechnology (EEC-1227110).